\documentclass[preprint,prd,aps]{revtex4}

\begin{document}
\begin{titlepage}
\title{\large \bf Further Investigation on Chiral Symmetry Breaking \\ 
in a Uniform External Magnetic Field} 

\author{\bf P. Jasinski and C. N. Leung}
\affiliation{Department of Physics and Astronomy, 
University of Delaware\\
Newark, DE 19716 \\}

\begin{abstract}

We study chiral symmetry breaking in QED when a uniform external magnetic 
field is present.  We calculate higher order corrections to the dynamically 
generated fermion mass and find them to be small.  In so doing we correct 
an error in the literature regarding the matrix structure of the fermion 
self-energy.

\end{abstract}
\maketitle
\end{titlepage}


In Ref.~\cite{LLNA} the effects of an external magnetic field on 
chiral symmetry breaking was studied using quantum electrodynamics 
as the model gauge field theory.  The Schwinger-Dyson equation for 
the fermion self-energy in the quenched, ladder approximation was 
expressed in terms of Ritus' representation~\cite{R} for the exact 
fermion propagator in a constant magnetic field and an approximate 
solution for the dynamically generated fermion mass was found.  The 
infrared fermion mass found in Ref.~\cite{LLNA} was consistent with 
that obtained from other approaches~\cite{GMS}.

In the present work we will examine how robust this approximate 
result is by carrying out the calculation to the next leading order 
of the approximation.  In the course of our study, we discover an 
inconsistency in Ref.~\cite{LLNA} regarding the form of the dynamical 
fermion mass matrix.  We obtain the correct matrix structure and show 
that, despite this inconsistency, the infrared dynamical fermion mass 
found in Ref.~\cite{LLNA} remains correct.

Let us begin with the Schwinger-Dyson equation for the fermion self-energy 
$\tilde{\Sigma}_A$ in the quenched, ladder approximation (see Eq.~(34) in 
Ref.~\cite{LLNA}): 
\begin{eqnarray}
\tilde{\Sigma}_A(\bar{p}) \delta_{kk'} 
&~=~&
i e^2 (2 |eH|) \sum_{k''=0}^{\infty} \sum_{\{\sigma\}} \int \frac
{d^4\hat{q}}{(2 \pi)^4} 
\frac{{\rm e}^{i {\rm sgn}(eH)(n-n''+\tilde{n}''-n')\varphi}}
{\sqrt{n!n'!n''!\tilde{n}''!}} 
\nonumber \\
& &
\cdot~ {\rm e}^{- \hat{q}_\perp^2} J_{nn''}(\hat{q}_\perp) 
J_{\tilde{n}''n'}(\hat{q}_\perp) \frac{1}{\hat{q}^2} 
\left(g_{\mu\nu} - (1 - \xi) \frac{\hat{q}_\mu \hat{q}_\nu}
{\hat{q}^2}\right) \nonumber \\
& &
\cdot~ \Delta(\sigma) \gamma^\mu \Delta(\sigma'') \frac{1}{\gamma \cdot 
\bar{p}'' + \tilde{\Sigma}_A(\bar{p}'')} \Delta(\tilde{\sigma}'') 
\gamma^\nu \Delta(\sigma').
\label{SD1}
\end{eqnarray}
We adopt the notation and convention of Ref.~\cite{LLNA} and will not 
explain them here.  Since we are interested in finding a solution for 
the infrared fermion mass, the calculation will be simplified by 
considering the $\bar{p} \rightarrow 0$ limit in Eq.~(\ref{SD1}).  Let 
us remind the reader that the 4-momentum $\bar{p}$ has the components 
$(p_0, 0, -{\rm sgn}(eH)\sqrt{2|eH|k}, p_3)$.  Consider first the $k 
\rightarrow 0$ limit.  Due to the presence of $\delta_{kk'}$ on the 
left hand side of Eq.~(\ref{SD1}), letting $k=0$ implies $k'=0$ which 
in turn requires $n = n' = 0$ and $\sigma = \sigma' = {\rm sgn}(eH)$.  
This is because the quantum numbers $n$, $k$, and $\sigma$ are related 
by the relation
\begin{equation}
n = n(k,\sigma) = k + \frac{\sigma}{2} {\rm sgn}(eH) - \frac{1}{2}
\label{index}
\end{equation}         
with the allowed values $n = 0, 1, 2, ...$, $k = 0, 1, 2, ...$, and 
$\sigma = \pm 1$, and similarly for $n' = n(k',\sigma')$.  In this limit, 
the $J$ polynomials in Eq.~(\ref{SD1}) become 
\begin{eqnarray}
 J_{0n''}(\hat{q}_{\perp})&=&[i~{\rm sgn}(eH) \hat{q}_{\perp}]^{n''},  
 \nonumber \\
 J_{\tilde{n}''0}(\hat{q}_{\perp})&=&[i~{\rm sgn}(eH) \hat{q}_
 {\perp}]^{\tilde{n}''}.
\end{eqnarray}
and Eq.~(\ref{SD1}) now reads
\begin{eqnarray}
 \tilde{\Sigma}_A(\bar{p}_\parallel) &~=~& ie^{2}(2|eH|)\sum_{k''=0}^
 {\infty} \sum_{\sigma'',\tilde{\sigma}''} \int\frac{d^4\hat{q}}{(2\pi)^4}
 \frac{{\rm e}^{i {\rm sgn}(eH)(\tilde{n}'' - n'')\varphi}}{\sqrt{n''!
 \tilde{n}''!}} {\rm e}^{-\hat{q}_{\perp}^2} [i~{\rm sgn}(eH) 
 \hat{q}_\perp]^{n''+\tilde{n}''} 
 \nonumber \\
 & & 
 \cdot~ \frac{1}{\hat{q}^2} \left(g_{\mu\nu} - 
 (1 - \xi) \frac{\hat{q}_\mu \hat{q}_\nu}{\hat{q}^2}\right) 
 \cdot~ \Delta\gamma^\mu \Delta(\sigma'') \frac{1}{\gamma \cdot
 \overline{p}'' + \tilde{\Sigma}_A(\overline{p}'')} \Delta(\tilde{\sigma}'')
 \gamma^\nu \Delta
\label{SD2}
\end{eqnarray}	
where $\Delta$ is now understood to be $\Delta = \Delta({\rm sgn}(eH))$, 
$\bar{p}_\parallel = (p_0, p_3)$, and the 4-momentum 
$\overline{p}''$ in the integrand is given by $\overline{p}''$~$=$~
$(p_0 - q_0$, $0$, $- {\rm sgn}(eH) \sqrt{2|eH|k''}$, $p_3 - q_3)$.  

By transforming to polar coordinates for the integrals over $\hat{q}_1$ 
and $\hat{q}_2$: $\int d\hat{q}_1 d\hat{q}_2 = \int \hat{q}_{\perp} 
d\hat{q}_{\perp} d\varphi$, the integration over the angle $\varphi$ can be 
carried out to yield 
\begin{equation}
 \int d\varphi {\rm e}^{i {\rm sgn}(eH)(\tilde{n}'' - n'')\phi} = 2 \pi 
 \delta_{\tilde{n}''n''} = 2 \pi \delta_{\tilde{\sigma}''\sigma''}.
\end{equation}
This allows us to rewrite Eq.~(\ref{SD2}) as
\begin{eqnarray}
 \tilde{\Sigma}_A(\bar{p}_\parallel) &~=~& ie^{2}(2|eH|) \sum_{k''=0}^{\infty} 
 \sum_{\sigma''} \frac{1}{n''!} \int\frac{d^{4}\hat{q}}{(2\pi)^4} 
 {\rm e}^{-\hat{q}_{\perp}^2} \frac{(-\hat{q}_\perp^2)^{n''}}{\hat{q}^2} 
 \nonumber \\
 & & 
 \cdot~ \left(g_{\mu\nu} - (1 - \xi) \frac{\hat{q}_\mu \hat{q}_\nu}{\hat{q}^2}
 \right) \Delta \gamma^\mu \Delta'' \frac{1}{\gamma \cdot \overline{p}'' + 
 \tilde{\Sigma}_A(\overline{p}'')} \Delta'' \gamma^\nu \Delta,
\label{SD3}
\end{eqnarray}	
where $\Delta''~=~\Delta(\sigma'')$.

The fermion self-energy is expected to have the form $\tilde{\Sigma}_A(\bar{p}) 
= Z(\bar{p}) \gamma \cdot \bar{p} + \Sigma_A(\bar{p})$, where 
$\Sigma_A(\bar{p})$ is a matrix representing the dynamically generated fermion 
mass.  An ansatz was made in Ref.~\cite{LLNA} that $\Sigma_A$ was proportional 
to the unit matrix.  An approximate solution for $\Sigma_{A}$ was then obtained 
by keeping only the dominant $k'' = 0$ term on the right hand side of 
Eq.~(\ref{SD3}).  In this case, $n''=0$ and $\sigma'' = {\rm sgn}(eH)$.  
However, in the calculation of Ref.~\cite{LLNA}, the spin summations in 
Eq.~(\ref{SD1}) were carried out before the infrared limit was taken (see 
Eqs.(44)-(47) there), and contributions from both $\sigma'' = {\rm sgn}(eH)$ 
and $\sigma'' = - {\rm sgn}(eH)$ were included for the $k'' = 0$ term, thus 
leading to the incorrect conclusion that $\Sigma_A$ was proportional to the 
unit matrix.  We shall see below that $\Sigma_A$ should be proportional to 
the matrix $\Delta({\rm sgn}(eH))$, which is equal to $diag (1,0,1,0)$ for 
${\rm sgn}(eH) = +1$ and equal to $diag (0,1,0,1)$ for ${\rm sgn}(eH) = -1$.  

If we let $\Sigma_{A}(\bar{p}) = m(\bar{p}) \Delta$, where $m(\bar{p})$ 
denotes the dynamically generated fermion mass, and keep only the $k''=0$ 
contributions, Eq.~(\ref{SD3}) becomes 
\begin{eqnarray}
 Z(\bar{p}_\parallel) \gamma \cdot \bar{p}_\parallel + m(\bar{p}_\parallel) 
 \Delta &~\simeq~& ie^{2}(2|eH|) \int\frac{d^{4}\hat{q}}{(2\pi)^4} 
 \frac{{\rm e}^{-\hat{q}_{\perp}^2}}{\hat{q}^2} \left(g_{\mu\nu} - (1 - \xi) 
 \frac{\hat{q}_\mu \hat{q}_\nu}{\hat{q}^2} \right) 
 \nonumber \\
 & & 
 \cdot~ \Delta \gamma^\mu \Delta \frac{1}{[1+Z(\overline{p}''_\parallel)] \gamma 
 \cdot \overline{p}''_\parallel + m(\overline{p}''_\parallel) \Delta} \Delta 
 \gamma^\nu \Delta,
\label{SD4}
\end{eqnarray}		
where $\overline{p}''_\parallel = \bar{p}_\parallel - q_\parallel$.  Because 
$\gamma \cdot \overline{p}''_\parallel$ commutes with $\Delta$, $(- [1+Z] 
\gamma \cdot \overline{p}''_\parallel + m \Delta)([1+Z] \gamma \cdot 
\overline{p}''_\parallel + m \Delta) = [1+Z]^2 (\overline{p}''_\parallel)^2 + 
m^2 \Delta$, which is a diagonal matrix that also commutes with $\gamma^\mu_
{\parallel} = (\gamma^0, \gamma^3)$.  We can therefore rewrite Eq.~(\ref{SD4}) 
as 
\begin{eqnarray}
 Z(\bar{p}_\parallel) \gamma \cdot \bar{p}_\parallel + m(\bar{p}_\parallel) 
 \Delta &~\simeq~& ie^{2}(2|eH|) \int\frac{d^{4}\hat{q}}{(2\pi)^4} 
 \frac{{\rm e}^{-\hat{q}_{\perp}^2}}{\hat{q}^2} \left(g_{\mu\nu} - (1 - \xi) 
 \frac{\hat{q}_\mu \hat{q}_\nu}{\hat{q}^2} \right) 
 \nonumber \\
 & & 
 \cdot~ \Delta \gamma^\mu \Delta \frac{-[1+Z(\overline{p}''_\parallel)] \gamma \cdot 
 \overline{p}''_\parallel + m(\overline{p}''_\parallel) \Delta}{[1+Z(\overline{p}''
 _\parallel)]^2 (\overline{p}''_\parallel)^2 + m^2(\overline{p}''_\parallel) 
 \Delta} \Delta \gamma^\nu \Delta.
\label{SD5}
\end{eqnarray}		

If we work in the Feynman gauge ($\xi=1$) and use the property that 
\begin{equation}
 \Delta \gamma^\mu \Delta = \Delta \gamma^\mu_\parallel \Delta = 
 \gamma^\mu_\parallel \Delta = \Delta \gamma^\mu_\parallel,
\label{sameDelta}
\end{equation}
we find  
\begin{eqnarray}
 & & \Delta\gamma^\mu \Delta \frac{-[1+Z(\overline{p}''_\parallel)] \gamma \cdot 
 \overline{p}''_\parallel + m(\overline{p}''_\parallel) \Delta}{[1+Z(\overline{p}''
 _\parallel)]^2 (\overline{p}''_\parallel)^2 + m^2(\overline{p}''_\parallel) 
 \Delta} \Delta \gamma_\mu \Delta
\nonumber \\
 &=& \frac{- 2 m(\overline{p}''_\parallel) \Delta}{[1+Z(\overline{p}''
 _\parallel)]^2 (\overline{p}''_\parallel)^2 + m^2(\overline{p}''_\parallel) 
 \Delta}
\nonumber \\
 &=& \frac{- 2 m(\overline{p}''_\parallel)}{[1+Z(\overline{p}''
 _\parallel)]^2 (\overline{p}''_\parallel)^2 + m^2(\overline{p}''_\parallel)} 
 \Delta
\end{eqnarray}
and Eq.~(\ref{SD5}) becomes
\begin{equation}
 Z(\bar{p}_\parallel) \gamma \cdot \bar{p}_\parallel + m(\bar{p}_\parallel) 
 \Delta ~\simeq~ -ie^{2}(2|eH|) \int\frac{d^{4}\hat{q}}{(2\pi)^4} 
 \frac{{\rm e}^{-\hat{q}_{\perp}^2}}{\hat{q}^2} 
 \frac{2 m(\overline{p}''_\parallel)}{[1+Z(\overline{p}''_\parallel)]^2 
 (\overline{p}''_\parallel)^2 + m^2(\overline{p}''_\parallel)} \Delta.
\label{SD6}
\end{equation}
It follows that $Z(\bar{p}_\parallel)$ vanishes in the Feynman gauge, as was 
found in Ref.~\cite{LLNA}, and one obtains the Schwinger-Dyson equation for 
the dynamical fermion mass:
\begin{equation}
 m(\bar{p}_\parallel) \Delta ~\simeq~ -ie^{2}(2|eH|) \int\frac{d^{4}\hat{q}}
 {(2\pi)^4} \frac{{\rm e}^{-\hat{q}_{\perp}^2}}{\hat{q}^2} 
 \frac{2 m(\overline{p}''_\parallel)}{(\overline{p}''_\parallel)^2 + 
 m^2(\overline{p}''_\parallel)} \Delta.
\label{SD7}
\end{equation}
Note that both sides of this equation are proportional to the $\Delta$ 
matrix, which justifies our ansatz for the matrix structure of $\Sigma_A$.  

Following the authors of Ref.~\cite{LLNA}, we seek a solution for $m$ in 
the infrared limit ($\bar{p}_\parallel \rightarrow 0$) and approximate 
$m(\overline{p}''_\parallel)$ that appears in the integrand on the right 
hand side of Eq.~(\ref{SD7}) by its infrared value.  This leads to the 
gap equation
\begin{equation}
 1 \simeq e^{2}(4|eH|) \int\frac{d^{4}\hat{q}}{(2\pi)^4} 
 \frac{{\rm e}^{-\hat{q}_{\perp}^2}}{\hat{q}^2} 
 \frac{1}{2|eH| \hat{q}_\parallel^2 + m^2} 
\label{gapeq}
\end{equation} 
where we have made a Wick rotation to Euclidean space.  This is 
precisely the gap equation (52) in Ref.~\cite{LLNA}, which has 
a solution of the form
\begin{equation} 
 m_0 \simeq a \sqrt{|eH|}{\rm e}^{-b\sqrt{\frac{\pi}{\alpha}}} 
\label{m0}
\end{equation} 
where $a$ and $b$ are real positive constants of order one and $\alpha$ 
is the fine structure constant.  The subscript $0$ indicates that this 
is the lowest order approximation for $m$.  As noted in Ref.~\cite{LLNA}, 
this solution is applicable for small $\alpha$.

Our result confirms the finding for the lowest order solution to the 
Schwinger-Dyson equation obtained in Ref.~\cite{LLNA}.  We also obtain 
the correct matrix structure for $\Sigma_A$, which is proportional to 
the matrix $\Delta({\rm sgn}(eH))$.  

Let us consider now the higher order corrections for the dynamical 
fermion mass $m$.  In the Feynman gauge, $Z(\bar{p}_\parallel) = 0$ and 
the Schwinger-Dyson equation (\ref{SD3}) may be written in the infrared 
limit as 
\begin{eqnarray}
 m \Delta &=& m_0 \Delta + ie^{2}(2|eH|) \sum_{k''=1}^\infty 
 \sum_{\sigma''} \frac{1}{n''!} \int\frac{d^{4}\hat{q}}{(2\pi)^4} 
 {\rm e}^{-\hat{q}_{\perp}^2} \frac{(-\hat{q}_\perp^2)^{n''}}{\hat{q}^2}
 \nonumber \\
 & & 
 \cdot~ \Delta\gamma^\mu \Delta'' \frac{1}{\gamma\cdot\overline{p}'' 
 + m(\overline{p}'') \Delta} \Delta'' \gamma_\mu \Delta,
\label{SD8}
\end{eqnarray}	
where we have set $\Sigma_{A}(0) = m \Delta$ and separated the $k''=0$ term 
which contributes to the lowest order solution $m_0$.  We shall estimate 
the next order corrections to $m$ by replacing $m(\overline{p}'')$ in the 
integrand with the lowest order solution $m_0$.  We need therefore to 
evaluate 
\begin{eqnarray}
 (m - m_0) \Delta &\simeq& ie^{2}(2|eH|) \sum_{k''=1}^\infty 
 \sum_{\sigma''} \frac{1}{n''!} \int\frac{d^{4}\hat{q}}{(2\pi)^4} 
 {\rm e}^{-\hat{q}_{\perp}^2} \frac{(-\hat{q}_\perp^2)^{n''}}{\hat{q}^2}
 \nonumber \\
 & & 
 \cdot~ \Delta\gamma^\mu \Delta'' \frac{1}{\gamma\cdot\overline{p}'' 
 + m_0 \Delta} \Delta'' \gamma_\mu \Delta,
\label{SD9}
\end{eqnarray}	
where $\overline{p}''$~$=$~$(- q_0$, $0$, $- {\rm sgn}(eH) \sqrt{2|eH|k''}$, 
$- q_3)$.

We shall consider the corrections to $m$ coming from the following 
contributions.  First, for $n''=0$, there is a subdominant $k''=1$ 
(and hence $\sigma'' = - {\rm sgn}(eH)$) term which was not included 
in $m_0$.  Second, there are two terms for $n''=1$: one for $k''=1$ 
and $\sigma'' = {\rm sgn}(eH)$, the other for $k''=2$ and $\sigma'' 
= - {\rm sgn}(eH)$.

For $k'' \neq 0$, $(\gamma\cdot\overline{p}'')$ no longer commutes 
with $\Delta$, and the expression for $(\gamma\cdot\overline{p}'' 
+ m_0 \Delta)^{-1}$ is no longer so simple and depends on 
${\rm sgn}(eH)$.  We shall present our calculation below for the 
case ${\rm sgn}(eH) = +1$.  However, the final result is applicable 
for either sign.  For ${\rm sgn}(eH) = +1$, one finds that 
$[\gamma\cdot\overline{p}'' + m_0 \Delta(1)]^{-1}$ is given by  
\begin{displaymath}
 \frac{1}{(\lambda \beta - \kappa^2)^2 - \lambda \beta m_0^2} \left(
 \begin{array}{cccc}
 -\lambda \beta m_0 & i \kappa \beta m_0 & \beta(\lambda \beta - \kappa^2) 
 & i \kappa(\lambda \beta - \kappa^2) \\ 
 i \kappa \lambda m_0 & \kappa^2 m_0 & -i \kappa(\lambda \beta - \kappa^2) 
 & \lambda(\lambda \beta - \kappa^2 - m_0^2) \\ 
 \lambda(\lambda \beta -\kappa^2) & -i \kappa(\lambda \beta - \kappa^2) 
 & -\lambda \beta m_0 & -i \kappa \lambda m_0 \\
 i \kappa(\lambda \beta - \kappa^2) & \beta(\lambda \beta - \kappa^2 - m_0^2) 
 & -i \kappa \beta m_0 & \kappa^2 m_o 
\end{array} \right) 
\end{displaymath}
where $\kappa \equiv \sqrt{2|eH|k''}$, $\lambda \equiv q_0+q_3$, and 
$\beta \equiv q_0-q_3$.  Fortunately, this simplifies a great deal after 
we compute $\Delta \gamma^\mu \Delta'' (\gamma\cdot\overline{p}'' + m_0 
\Delta)^{-1} \Delta'' \gamma_\mu \Delta$.

Since $\sigma'' = \pm 1$, and 
\begin{equation}
 \Delta(1) \gamma^j_\perp \Delta(1) = 0 = \Delta(1) 
 \gamma^\mu_\parallel \Delta(-1), 
\end{equation}
where $\gamma^j_\perp = (\gamma^1, \gamma^2)$, we need only evaluate 
\begin{eqnarray}
 &&\Delta(1) \gamma^\mu_\parallel \Delta(1) [\gamma\cdot\overline{p}'' 
 + m_0 \Delta(1)]^{-1} \Delta(1) \gamma_{\parallel \mu} \Delta(1) 
 \nonumber \\
 &=& \gamma^\mu_\parallel \Delta(1) [\gamma\cdot\overline{p}'' 
 + m_0 \Delta(1)]^{-1} \Delta(1) \gamma_{\parallel \mu} 
 \nonumber \\
 &=& \frac{2\lambda \beta m_0}{(\lambda \beta - \kappa^2)^2 - \lambda 
 \beta m_0^2} \cdot \Delta(1)
\label{matrix+1}
\end{eqnarray}
and 
\begin{eqnarray}
 &&\Delta(1) \gamma^j_\perp \Delta(-1) [\gamma\cdot\overline{p}'' 
 + m_0 \Delta(1)]^{-1} \Delta(-1) \gamma_{\perp j} \Delta(1) 
 \nonumber \\
 &=& \gamma^j_\perp \Delta(-1) [\gamma\cdot\overline{p}'' 
 + m_0 \Delta(1)]^{-1} \Delta(-1) \gamma_{\perp j} 
 \nonumber \\
 &=& \frac{-2 \kappa^2 m_0}{(\lambda \beta - \kappa^2)^2 - \lambda 
 \beta m_0^2} \cdot \Delta(1) 
 \nonumber \\
 & & +~ \frac{2(\lambda \beta - \kappa^2 - m_0^2)}{(\lambda \beta - 
 \kappa^2)^2 - \lambda \beta m_0^2}
 \left(
 \begin{array}{cc}
 0 & \frac{\beta}{2}(1 + \sigma_3) \\ 
 \frac{\lambda}{2}(1 + \sigma_3) & 0 
\end{array} \right) 
\label{matrix-1}
\end{eqnarray}
where $\sigma_3$ denotes the Pauli matrix.  The off-diagonal elements 
in Eq.~(\ref{matrix-1}) might seem to be a problem.  However, more 
careful inspection shows that they are odd in $q_0$ and $q_3$ and will 
vanish upon integration over these variables, thus reducing the matrix 
to a diagonal matrix proportional to $\Delta(1)$.  It should be stressed 
that both matrix structures appearing in Eqs.(\ref{matrix+1}) and 
(\ref{matrix-1}) are proportional to $\Delta(1)$, consistent with the 
left-hand-side of Eq.~(\ref{SD9}).  This shows that $\Sigma_A$ is 
proportional to the matrix $\Delta({\rm sgn}(eH))$ for all higher order 
terms in Eq.~(\ref{SD8}).

Putting all the pieces together, using Eq.~(\ref{matrix+1}) for the 
$k''=1$, $\sigma'' = 1$ term and Eq.~(\ref{matrix-1}) for the other 
two terms corresponding to $\sigma'' = -1$ and $k''=1,2$, we find the next 
order corrections for the dynamical fermion mass to be
\begin{eqnarray}
 (m - m_0) \Delta &\simeq& ie^{2}(2|eH|) \int\frac{d^{4}\hat{q}}{(2\pi)^4} 
 \frac{{\rm e}^{-\hat{q}_{\perp}^2}}{\hat{q}^2} \left\{\frac{-4|eH|m_0}
 {(-q_0^2+q_3^2+2|eH|)^2 + m_0^2(-q_0^2+q_3^2)} \right.
 \nonumber \\
 &~& + \hat{q}_\perp^2 \left[\frac{2m_0(-q_0^2+q_3^2)}
 {(-q_0^2+q_3^2+2|eH|)^2 + m_0^2(-q_0^2+q_3^2)} \right. 
 \nonumber \\
 &~& + \left. \left. \frac{8|eH|m_0}{(-q_0^2+q_3^2+4|eH|)^2 + 
 m_0^2(-q_0^2+q_3^2)} \right] \right\} \Delta.
\label{deltam1}
\end{eqnarray}
In other words, 
\begin{eqnarray}
 \frac{m - m_0}{m_0} &\simeq& \frac{\alpha}{\pi}|eH| \int_0^\infty 
 d\hat{q}_\perp^2 \int_0^\infty d\hat{q}_{\parallel}^2 
 \frac{e^{-\hat{q}_{\perp}^2}}{\hat{q}_\parallel^2+\hat{q}_\perp^2} 
 \left\{\frac{1}{2|eH|(1+\hat{q}_\parallel^2)^2 + 
 m_0^2 \hat{q}_\parallel^2} \right.
 \nonumber \\
 &~& - \left. \hat{q}_\perp^2 \left[\frac{\hat{q}_\parallel^2}
 {2|eH|(1+\hat{q}_\parallel^2)^2 + m_0^2 \hat{q}_\parallel^2}
 + \frac{2}{2|eH|(2+\hat{q}_\parallel^2)^2 + m_0^2 
 \hat{q}_\parallel^2} \right] \right\}
\label{deltam2}
\end{eqnarray}
where we have performed a Wick rotation to Euclidean space and 
carried out the integration over the polar angles on the 
$\hat{q}_\parallel$ plane as well as on the $\hat{q}_\perp$ plane. 

We see that the fractional correction to $m$ is quite small, being 
proportional to $(\alpha/\pi)$.  If we substitute the expression 
for $m_0$ found in Eq.~(\ref{m0}), with $a$ and $b$ set to be one, 
we can evaluate the integrals numerically to obtain the estimate 
\begin{eqnarray}
 \frac{m - m_0}{m_0} &~\simeq~& \left(\frac{\alpha}{\pi}\right) 
 (0.4275 - 0.2161 - 0.1628) \nonumber \\
 &~\simeq~& 0.0486 \left(\frac{\alpha}{\pi}\right), 
\end{eqnarray}
a very small number indeed.

In summary, we have reexamined the Schwinger-Dyson equation analysis 
of chiral symmetry breaking in QED in the presence of a uniform 
external magnetic field, confirming the lowest order result for 
the dynamical fermion mass obtained in Ref.~\cite{LLNA}, and finding 
the correct matrix structure for the fermion self-energy.  We have 
also calculated the higher order corrections to the dynamical fermion 
mass and found them to be small, thus validating the approximation 
scheme used in the calculation.

\newpage

\begin{center} 
{\bf ACKNOWLEDGEMENTS}\\
\end{center}

This work was supported in part by the U.S. Department of Energy under 
Grant No. DE-FG02-84ER40163.  Part of this work was carried out when 
C.N.L. was visiting the Academia Sinica in Taipei, Taiwan.  He would 
like to thank members of the Institute of Physics there for their warm 
hospitality.  
\raggedbottom

\end{document}